\documentstyle[aps,pre,epsfig,multicol]{revtex}

\def\bneta{\mbox{\boldmath $\eta$}}
\def\bzeta{\mbox{\boldmath $\zeta$}}
\def\bxi{\mbox{\boldmath $\xi$}}
\def\sign{\mbox{\rm sign}}

\begin{document}
\draft
\title{Optimal coloured perceptrons}
\author{ D.~Boll\'e\thanks{E-mail:
	   desire.bolle@fys.kuleuven.ac.be.}
           \thanks{Also at Interdisciplinair Centrum
            voor Neurale Netwerken, K.U.Leuven, Belgium.}
	    and
         P.~Koz{\l}owski
	  \footnotemark[2]
           \thanks{E-mail:
	   piotr.kozlowski@fys.kuleuven.ac.be}
            }
\address{Instituut voor Theoretische Fysica,
            Katholieke Universiteit Leuven, \\ B-3001 Leuven, Belgium}

\maketitle
\begin{abstract}
Ashkin-Teller type perceptron models are introduced. Their maximal
capacity per number of couplings is calculated within a first-step
replica-symmetry-breaking Gardner approach. The results are compared
with extensive numerical simulations using several algorithms. 
\end{abstract}

\begin{multicols}{2}
\narrowtext

\section{Introduction}
The perceptron which was first analyzed with statistical mechanics
techniques in the seminal paper of Gardner \cite{Ga88} is by now a
well-known  
and standard model in theoretical studies and practical applications in 
connection with learning and generalization \cite{HKP91,MRS95,OK96,Sa98}. 
A number of extensions of the perceptron model have been formulated,    
including many-state and graded-response perceptrons (e.g.,
\cite{NR91,GK94,Ko90,MKB91,BDM91,BKM93}). Here we present some new
extensions allowing for so-called coloured or Ashkin-Teller type
neurons, i.e., different types of binary neurons at each site possibly
having  different functions. 

The idea of looking at such a model is based upon our recent work on 
Ashkin-Teller recurrent neural networks \cite{BK1,BK2}. There we
showed that for this model with two types of binary neurons interacting
through a four-neuron term and equipped with a Hebb learning rule, both
the thermodynamic and dynamic properties suggest that such a model can
be more efficient than a sum of two Hopfield models. For example, the
quality of pattern retrieval is enhanced through a larger overlap at higher
temperatures and the maximal capacity is increased. For more details
and an underlying neurobiological motivation for the introduction of
different types of neurons we refer to \cite{BK2}.

In the light of these results an interesting question is whether such a
coloured perceptron can still be more efficient than the standard
perceptron. In other words, can it have a larger maximal capacity than
the one of a standard perceptron, which is known \cite{Ga88} to be
$\alpha_c=2$ (for random uncorrelated patterns). It has been
suggested that this number is characteristic for all binary networks 
independent of the multiplicity of the neuron interactions.
Thereby, the capacity is defined as the thermodynamic limit 
of the ratio of the total number of bits per (input) neuron to be stored
and  the total number of couplings  
per (output) neuron \cite{Ko90}. We remark that "input" and "output" refer
specifically to the perceptron case.

In the sequel the maximal capacity of coloured perceptron models is
studied using the Gardner approach \cite{Ga88,GD88}. First-step         
replica-symmetry-breaking effects are evaluated and the analytic 
results are compared with extensive numerical simulations
using  various learning algorithms. 

The rest of this paper is organized as follows. In Sec.\ II we introduce
two Ashkin-Teller type perceptron models. Section III contains the
replica theory and determines the maximal capacity by calculating the
available volume in the space of couplings both in the replica-symmetric
(Sec.\ IIIA) and the first-step replica-symmetry-breaking approximation
(Sec.\ IIIB). Section IV describes the results of numerical simulations
with algorithms obtained by generalizing various algorithms for simple
perceptrons. In Sec.\ V we present our conclusions. Finally, two
appendices contain some technical details of the derivations.
 
\section{The model}

Let us first formulate the coloured perceptron models.
We consider $p$ input patterns 
$\bzeta^\mu=\{\zeta_i^\mu\}=\{\xi_i^\mu,\eta_i^\mu\},~
i=1,\ldots,N$ consisting out of two different types of patterns
$\bxi^\mu=\{\xi_i^\mu\}$ and $\bneta^\mu=\{\eta_i^\mu\}$,
and a corresponding set of outputs $\bzeta_0^\mu =
\{\xi_0^\mu,\eta_0^\mu\}~\mu=1,\ldots,p$ which are determined by
\begin{eqnarray}
       \xi_0^\mu&=&\sign(h_1^\mu +\eta_0^\mu h_3^\mu )\label{e.s1}\\
        \eta_0^\mu&=&\sign(h_2^\mu +\xi_0^\mu h_3^\mu )\label{e.s2}\\
	\xi_0^\mu\eta_0^\mu&=&\sign(\eta_0^\mu h_1^\mu +\xi_0^\mu h_2^\mu)
	\label{e.s3}
\end{eqnarray}
\noindent where $h_r$ ($r=1,2,3$) are the local fields acting on the patterns
$\xi$, $\eta$ and their product $\xi\eta$ respectively
\begin{eqnarray}
&&  h_1^\mu = \frac{1}{n_1} \sum_i J_i^{(1)} \xi_i^\mu,~~~~ 
  h_2^\mu = \frac{1}{n_2} \sum_i J_i^{(2)} \eta_i^\mu,~~~~\\
&&  h_3^\mu = \frac{1}{n_3} \sum_i J_i^{(3)} \xi_i^\mu \eta_i^\mu,~~~
n_r^2 = \sum_i(J_i^{(r)})^2,~~r=1,2,3\,.
     \label{eq:field}
\end{eqnarray}
Both types of input patterns and their corresponding outputs are supposed
to be independent identically
distributed random variables (IIDRV) taking the values $+1$ or $-1$
with probability $1/2$. 
The set of three equations (\ref{e.s1})-(\ref{e.s3}) defines a
mapping of the inputs $\bzeta_i^\mu$ onto the corresponding outputs 
$\bzeta_0^\mu$. We call it model I. We remark that the
specific form of the equations (\ref{e.s1})-(\ref{e.s3}) is related to
the transition probabilities for a spin-flip in the dynamics \cite{BK1}.
A second model, denoted by II, is defined by considering only the two
equations (\ref{e.s1}) and (\ref{e.s2}). When $|h_3|>|h_1|$ and
$|h_3|>|h_2|$ then the relations (\ref{e.s1})-(\ref{e.s2}) are satisfied
by two (out of the four possible) values of the output $\bzeta_0$, 
otherwise model II gives the same output as model I. In other words, due
to the presence of the $\eta_0^\mu$ and $\xi_0^\mu$ in the gain
functions, model II contains more freedom and, strictly speaking, it is
not a mapping. 

The sequential dynamics of these two models has been 
studied in the case of low loading with the Hebb
rule and shown to lead to the same equilibrium behaviour \cite{BK1}.
However, this is not guaranteed here since we are concerned with   
optimal couplings maximizing the loading capacity.  
At this point we remark that when all $J^{(3)}_i$ are equal to zero we find
back two independent standard binary perceptron models.
In the sequel we take the couplings to satisfy the spherical constraint 
$n_r=\sqrt N$.

\section{Replica theory for the maximal capacity}

The coloured perceptron is trained to store correctly 
$p= \frac{3}{2}\alpha N$ patterns with $\alpha$  the
loading capacity. The factor ${3}/{2}$ follows naturally from the
definition of capacity given in the introduction. A pattern is stored   
correctly when the so-called  aligning field \cite{whyt} is bigger than
a certain
constant $\kappa \geq 0$ whereby the latter indicates the stability. It
is a measure for the size of the basin of attraction of that pattern. 
Specifically we require that
\begin{eqnarray}
    \lambda_{\xi}^\mu (\{J\})&=&
           \xi_0^\mu \left(h_1^\mu
                  +  \eta_0^\mu h_3^\mu\right)
                        > \kappa_\xi \geq 0 \label{eq:afieldxi} \\
     \lambda_{\eta}^\mu (\{J\})&=&
           \eta_0^\mu \left(h_2^\mu
                  +  \xi_0^\mu h_3^\mu\right)
                        > \kappa_\eta \geq 0 \label{eq:afieldeta} \\
     \lambda_{\xi\eta}^\mu (\{J\})&=&
            (\xi_0^\mu h_1^\mu
                  +  \eta_0^\mu h_2^\mu)
                          > \kappa_{\xi\eta} \geq 0 \,, 
       \label{eq:afieldxieta}
\end{eqnarray}
with $\{J\}= \{J_i^{(r)}\}$ denoting the configurations in
the space of interactions. 
For $\kappa_\xi=\kappa_\eta=\kappa_{\xi\eta}=0$ all patterns that
satisfy equations (\ref{e.s1})-(\ref{e.s3}) also
satisfy (\ref{eq:afieldxi})-(\ref{eq:afieldxieta}).
We remark that for model~II the last inequality is superfluous.
\noindent\rule{20.5pc}{0.1mm}\rule{0.1mm}{1.5mm}\hfill

The aim is then to determine the maximal value of the loading $\alpha$ 
for which couplings satisfying (\ref{eq:afieldxi})-(\ref{eq:afieldxieta})
can still be found.
In particular, the question whether this model can be more efficient
than the existing two-state models is relevant. 
 
Following refs.~\cite{Ga88,GD88} we formulate the problem as an energy
minimization in the space of couplings with the formal energy function  
defined as
\begin{eqnarray}
          E(\{J\})&=&\sum_\mu \left[1\right.-
                  \Theta(\lambda_\xi^\mu(\{J\}) - \kappa_\xi)\nonumber \\
       &\times&
               \left.\Theta(\lambda_\eta^\mu(\{J\}) - \kappa_\eta)
       \Theta(\lambda_{\xi\eta}^\mu(\{J\}) -\kappa_{\xi\eta})\right]\,.
    \label{eq:energyI}
\end{eqnarray}
We remark that for model~II the third $\Theta$-factor is absent. 
The quantity above counts the number of weakly embedded patterns, i.e.,
the patterns  with stability less than
$\kappa_\xi, \kappa_\eta, \kappa_{\xi\eta}$. Therefore, the minimal
energy  gives the minimal number of patterns that are stored
incorrectly. This number is zero below a maximal storage capacity
$\alpha_c(\kappa_\xi,\kappa_\eta, \kappa_{\xi\eta})$.

The basic quantity to start from is the partition function
\begin{eqnarray}
  Z(\beta)&=& 
  \langle \exp\left[ -\beta E(\{J\} )\right] \rangle_{\{J\}}\\
                      \langle ...\rangle_{\{J\}}
	 &=&\int \prod_i dJ_i
     \prod _r \delta \left(\sum_i (J_i^{(r)})^2 - N \right)...~,
  \end{eqnarray}
with $\beta$ the inverse temperature.
As usual it is $\ln Z$ which is assumed to be a self-averaging extensive
quantity \cite{Ga88,whyt}. The related free energy per site
\begin{equation}
  f = - \lim_{N \rightarrow \infty} \frac{1}{N\beta} \ln Z(\beta)
  \label{eq:minfraction}
\end{equation}
is equal, in the limit $\beta\rightarrow\infty$, to 
\begin{equation}
 \frac{\langle E\rangle_\infty}{N}
    \equiv \lim_{\beta\rightarrow\infty}
             \frac{\langle E(\{J\})\exp\left[ 
        -\beta E(\{J\})\right]\rangle_{\{J\}}} {NZ(\beta)} \,,
	\label{eq:flimit} 
\end{equation}
which is 
the minimal fraction of wrong patterns (recall eq. (\ref{eq:energyI})).

In order to perform the average over the disorder in the input patterns 
$\bzeta^\mu$ and the corresponding outputs $\zeta_0^\mu$ 
we employ the replica method.
The calculations proceed in a standard way although the technical
details are much  more complex. Introducing the order parameters
$q_{\gamma\tau}^{(r)} = \frac{1}{N}\sum_i J_i^{(r)\gamma}
J_i^{(r)\tau}$, with $r=1,2,3$ and $\gamma,\tau=1,...,n$ we write following  \cite{Ga88}

\end{multicols}
\widetext
\begin{eqnarray}
  \langle Z^n(\beta)\rangle
      &=&\int\prod_{r,\gamma,\tau>\gamma}\left(\frac{{\rm d}
        q_{\gamma\tau}^{(r)} {\rm d}\phi_{\gamma\tau}^r}{2\pi/N}\right)
       \prod_{r,\gamma}\frac{{\rm d}\epsilon_\gamma^r}{2 \pi} 
    \exp N\left\{\frac{3}{2}\alpha
       G_0(q_{\gamma\tau}^{(r)})
       +\sum_{r,\gamma,\tau>\gamma}iq_{\gamma\tau}^{(r)}
        \phi_{\gamma\tau}^r
      +G_1(\phi_{\gamma\tau}^r,\epsilon_\gamma^r)\right\}\label{zpn} 
                       \\ 
     G_0&=&\ln\left\{\prod_\gamma\left(\left[e^{-\beta}
                \int\prod_{r'}\frac{{\rm d}\lambda^{r'\gamma}}{2\pi}
     +\left(1-e^{-\beta}\right)
    \int_{\kappa_\xi}^\infty\int_{\kappa_\eta}^\infty
             \int_{\kappa_{\xi\eta}}^\infty
     \prod_{r'}\frac{{\rm d}\lambda^{r'\gamma}}{2\pi}\right]
        \int\prod_{r'} {\rm d}x^{r'\gamma}
           \exp\left\{\sum_{r'} ix^{r'\gamma}\lambda^{r'\gamma}
                   \right.\right.\right.
	          \nonumber\\
     &-& \frac{1}{2}\left[\left(x^{1\gamma}+x^{3\gamma}\right)^2	  
		  +\left(x^{2\gamma}+x^{3\gamma}\right)^2
            +\left(x^{1\gamma}+x^{2\gamma}\right)^2\right]
	           \nonumber\\    
     &-&\sum_{\tau>\gamma}\left[\left(x^{1\gamma}+x^{3\gamma}\right)
        \left(x^{1\tau}+x^{3\tau}\right)q_{\gamma\tau}^{(1)}
         +\left(x^{2\gamma}+x^{3\gamma}\right)
          \left(x^{2\tau}+x^{3\tau}\right)q_{\gamma\tau}^{(2)}\right.
              +\left.\left.\left.\left.\left(x^{1\gamma}+x^{2\gamma}\right)
        \left(x^{1\tau}+x^{2\tau}\right)q_{\gamma\tau}^{(3)}
	  \right]
	  \rule{0cm}{0.5cm}\right\}
	  \rule{0cm}{0.6cm}\right)
	  \rule{0cm}{0.7cm}\right\}\nonumber
       \\    
  G_1&=&\ln\left\{\int\prod_{r,\gamma}\left({\rm d} J^{(r)\gamma}\right)
   \exp\left[i\sum_{r,\gamma}\epsilon_\gamma^r((J^{(r)\gamma})^2-1)-
   i\sum_{r,\gamma,\tau >\gamma}\phi_{\gamma\tau}^r J^{(r)\gamma}J^{(r)\tau}
   \right]\right\},\nonumber
\end{eqnarray}

\begin{multicols}{2}
\narrowtext
\noindent where $\langle ...\rangle$ denotes the average over the patterns,
$r'=1,2,3$ for model~I and $1,2$ for model~II.
Because of the latter we 
remark that for model~II the formula for $G_0$ can be simplified: 
the integrals with respect to  $\lambda^{3\gamma}$ and
$x^{3\gamma}$ are not present and thus $x^{3\gamma}$, $x^{3\tau}$ 
and $\lambda^{3\gamma}$ have to be
set to zero. Because of this simplification we only outline explicitly
the calculations for model~II in the sequel. The corresponding formulas
for model~I can be found in Appendix~B.  

\subsection{Replica symmetric anzatz}

We continue by making the replica-symmetric (RS) anzatz 
$q_{\gamma\tau}^{(r)}= q^{(r)},\phi_{\gamma \tau}^{r}= i\phi^r, 
\epsilon_{\gamma}^{r}= i\epsilon^{r}$.
Moreover, for convenience, we set $q^{(1)}=q^{(2)}=q^{(3)}=q$. The
latter is justified for model~I because of the symmetry present in this
model. Furthermore, since we are going to take all $q^{(r)} \to 1$ in the
Gardner-Derrida analysis anyway, we keep this equality also for model~II. 
Taking then the limits 
$\beta\rightarrow\infty$, $N\rightarrow\infty$ and $n\rightarrow 0$ we  
arrive, in the case of model~II, at 

\begin{eqnarray}
v&=&\lim_{N\rightarrow\infty} \frac{1}{N} <\ln Z>\nonumber\\
 &=&\frac{3}{2}\alpha\int {\rm D}(s_1\left({q}/{2}\right))
  {\rm D}(s_2\left({3q}/{2}\right))
    \ln \psi_{RS}(\kappa_\xi,\kappa_\eta,s_1,s_2,q)\nonumber\\
   &+&\frac{3}{2}\left( \ln (1-q)+\frac{1}{1-q}+\ln 2\pi\right)
 \label{eq:result}
\end{eqnarray}
with 
\begin{equation}
\psi_{RS}(\kappa_\xi,\kappa_\eta,s_1,s_2,q)=
   \int_{l_1}^{\infty}\int_{l_2}^{l_3}
       \prod_{\nu}{\rm D}(s_{\nu}(1))
\end{equation}
where $\nu=1,2$, ${\rm D}(s(y)) ={\rm d}s\exp(-\frac{1}{2y}s^2)/\sqrt{2\pi y}$ is a
modified Gaussian measure,
\begin{eqnarray}
  l_1&=&\frac{\sqrt{\frac{2}{3}}(\frac{1}{2}(\kappa_\xi+\kappa_\eta)-s_2)}
		 {\sqrt{1-q}}~,\\
  l_2&=&\frac{\sqrt{2}(\kappa_\eta-s_2-s_1)}{\sqrt{1-q}}-u_2\sqrt{3}~,\\
  l_3&=&\frac{\sqrt{2}(-\kappa_\xi+s_2-s_1)}{\sqrt{1-q}}+u_2\sqrt{3}~.
\end{eqnarray}
and $q$ takes those values that minimize $v$, the available 

\noindent\hfill\rule[-1.5mm]{0.1mm}{1.5mm}\rule{20.5pc}{0.1mm}
volume in the space of couplings. 
For the corresponding expression in the case of model~I we refer to
Appendix~B.

Taking $\kappa_\xi=\kappa_\eta=\kappa$ and
supposing that the maximal capacity, $\alpha_c=\alpha_{RS}$, is signaled by
the Gardner-like criterion $q \rightarrow 1$ we obtain
\begin{eqnarray}
\alpha_{RS}(\kappa)=\hspace{7cm}\nonumber\\
\lim_{q\rightarrow 1}\left\{
\frac{-\ln (1-q)-\frac{1}{1-q}-\ln 2\pi}
{\int {\rm D}(s_1\left({q}/{2}\right))
  {\rm D}(s_2\left({3q}/{2}\right))
    \ln \psi_{RS}(\kappa,\kappa,s_1,s_2,q)}\right\}. 
\end{eqnarray}
This maximal capacity as a function of $\kappa$ is shown for both models 
in  figs.~\ref{res1} and \ref{res2} as a full line.
For model~I we obtain, e.g., $\alpha_{RS}(\kappa=0)=1.92$, a value that
is smaller than the Gardner capacity for the simple perceptron. For
model~II however, we get the interesting result that 
$\alpha_{RS}(\kappa=0)=2.74>2$.

\subsection{First-step replica symmetry breaking}

It is straightforward to show geometrically that learning almost
antiparallel patterns, i.e., patterns satisfying 
$(\bxi^\mu \xi_0^\mu,\bneta^\mu \eta_0^\mu) 
\approx -(\bxi^\nu \xi_0^\nu,\bneta^\nu \eta_0^\nu)$
results in a splitting of the space of couplings into disconnected
regions. This 
suggests that RS is broken and, consequently, the results for
$\alpha_{RS}$ found in Sec.\ IIIA are only upperbounds for the true 
capacity. Therefore, we want to improve the RS results by applying the
first step of Parisi's replica-symmetry-breaking (RSB) scheme (e.g.,
\cite{parisi}). So, we assume that the $q_{\gamma\tau}^{(r)}$ in  
equation (\ref{zpn}) have the following matrix block structure
\begin{equation}
q_{\gamma\tau}^{(r)}=\cases{
       q_1^{(r)}~~{\rm if}~
             {\rm int}\left(\frac{(\gamma-1)m}{n}\right)={\rm
                   int}\left(\frac{(\tau-1)m}{n}\right)\cr
       q_0^{(r)}~~{\rm otherwise}~,}\label{rsb1a}
\end{equation}
where $n$ is the size of the matrix $q_{\gamma\tau}^{(r)}$, $m$ is the
number of diagonal blocks and int($x$) denotes the integer part of $x$.

For model II we take $q_{\gamma\tau}^{(1)}=q_{\gamma\tau}^{(2)}\not=
q_{\gamma\tau}^{(3)}$ reflecting the symmetry of this model. For
model~I we repeat that all $q^{(r)}$'s can be taken equal. 
We then consider the limits $q_1^{(r)}\rightarrow 1$ and $n\rightarrow 0$ 
in such a way that $m/(1-q_1)$, with $q_1^{(1)}=q_1^{(2)}=q_1^{(3)}=q_1$,
remains finite. After a tedious calculation we arrive at the following 
expression for the RSB1 maximal capacity for model~II
\end{multicols}
\widetext

\begin{eqnarray}
 \alpha_{RSB1}(\kappa)=
     \min_{q_0^{(1)},q_0^{(3)},M}
     \left\{\frac{-\frac{2}{3}\left(\ln(1+M)
              +\frac{q_0^{(1)}M}{(1+M)(1-q_0^{(1)})}
	      +\frac{1}{2}\ln(1+M_3)+
	      \frac{1}{2}\frac{q_0^{(3)}M}{(1+M_3)(1-q_0^{(1)})}\right)}
     {\int {\rm D}t_1{\rm D}t_2\ln 
         \psi_{RSB1}\left(\kappa,t_1,t_2,q_0^{(1)},q_0^{(3)},M\right)}
                   \right\}
	\label{alphaRSB1}
\end{eqnarray}
\begin{multicols}{2}
\narrowtext
with 
\begin{eqnarray}
r_3=\frac{1-q_0^{(3)}}{1-q_0^{(1)}},~M_3=M r_3,~M=\frac{m(1-q_0^{(1)})}{1-q_1} 
\end{eqnarray}
and D$t_i={\rm d}t_i\exp(-\frac{1}{2}t_i^2)/\sqrt{2\pi}$ a Gaussian measure.
The explicit form of the function 
$\psi_{RSB1}(\kappa,t_1,t_2,q_0^{(1)},q_0^{(3)},M)$ 
can be found in appendix~A. An
analogous form for model~I is written down in Appendix~B.

The results are presented in figs. \ref{res1} and \ref{res2} as full
lines. As expected they lie below the RS results confirming the breaking
of RS, e.g., $\alpha_{RSB1}(\kappa=0)=1.83$ for model~I and $2.28$ for
model~II. We remark that the breaking for model~II is stronger than for
model~I, the reason being that model~II allows more freedom as explained
in the introduction.  Finally, on the basis of results in the literature
for the simple perceptron \cite{whyt}, \cite{BE99} we expect that the
RSB1 results are very close to the exact ones.
This is further examined by performing numerical simulations as described  
in the following section.

\section{Numerical simulations}

The idea of these simulations is to train the network with a certain
learning algorithm in order to learn as many random patterns as possible.
The main technical difficulties are to find an efficient algorithm and
prove its convergence.

We have tried to generalize various algorithms proposed for simple
perceptrons  \cite{KM,AK,ada,Im}. The most effective ones appeared to be
some particular generalization of the adaptive Gardner algorithm
\cite{AK} and the Adatron algorithm \cite{ada}. In the sequel we
only report on the results obtained with these two algorithms. We remark
that we have chosen $\kappa_\xi=\kappa_\eta=\kappa_{\xi\eta}=\kappa$ in
all simulations.

One of the algorithms that has demonstrated its efficiency and for which
convergence has been shown in the case of the standard perceptron is
given in ref.~\cite{AK}. It is an adaptive version of the original algorithm    
proposed by Gardner \cite{Ga88}.
Using heuristic arguments presented in \cite{AK} we have constructed for
the coloured perceptron model~II the following analogous learning rule
\begin{eqnarray}
   J_i^{(1)}&\rightarrow&J_i^{(1)}+\xi_0^{\mu}\xi_i^{\mu}\frac{1}{2}
                     \left(\kappa_\xi-\lambda_\xi^\mu\right)
                   \Theta\left(\kappa_\xi-\lambda_\xi^\mu\right)
		   \label{j1per}\\
    J_i^{(2)}&\rightarrow&J_i^{(2)}+\eta_0^{\mu}\eta_i^{\mu}\frac{1}{2}
                     \left(\kappa_\eta-\lambda_\eta^\mu\right)
                   \Theta\left(\kappa_\eta-\lambda_\eta^\mu\right)
		   \label{j2per}\\
    J_i^{(3)}&\rightarrow&J_i^{(3)}
                    +\xi_0^{\mu}\eta_0^{\mu}\xi_i^{\mu}\eta_i^{\mu}
            \frac{1}{2}\left[\left(\kappa_\xi-\lambda_\xi^\mu\right)
            \Theta\left(\kappa_\xi-\lambda_\xi^\mu\right)\right.\nonumber\\
          &+&\left(\kappa_\eta-\lambda_\eta^\mu\right)\left.
                \Theta\left(\kappa_\eta-\lambda_\eta^\mu\right)\right]
		\,.
\end{eqnarray}

\noindent\hfill\rule[-1.5mm]{0.1mm}{1.5mm}\rule{20.5pc}{0.1mm}
The form of the algorithm for model~I is a bit different and given in
Appendix~B.
This algorithm should be carried out sequentially over the patterns and
sequentially or parallel over the couplings as long as one of the
arguments of the $\Theta$ functions is positive.
It appears to have the characteristics of the most efficient, non-linear
algorithm discussed in \cite{AK}.

\begin{figure}[h]
\epsfysize=9.15cm
\epsfxsize=13cm
\centerline{\epsfbox{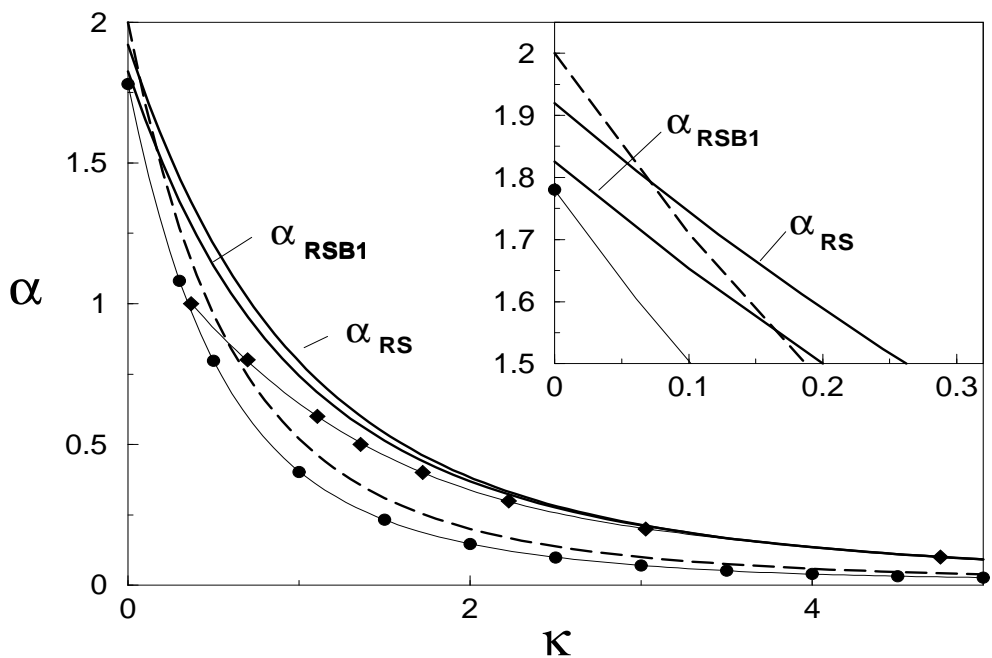}}
\caption{The maximal capacity of the coloured perceptron model I as a
function of $\kappa$. Theoretical results, $\alpha_{RS}$ and $\alpha_{RSB1}$,
versus simulations. 
The circles are the results for the adaptive Gardner algorithm, the
diamonds for the Adatron algorithm. The error bars are smaller than the
size  of the symbols (not in the inset). The solid thin lines are polynomial fits
to these results. The maximal capacity of a simple perceptron is
indicated with a broken line.}\label{res1}
\end{figure}

\begin{figure}[h]
\epsfysize=9.15cm
\epsfxsize=12.45cm
\centerline{\epsfbox{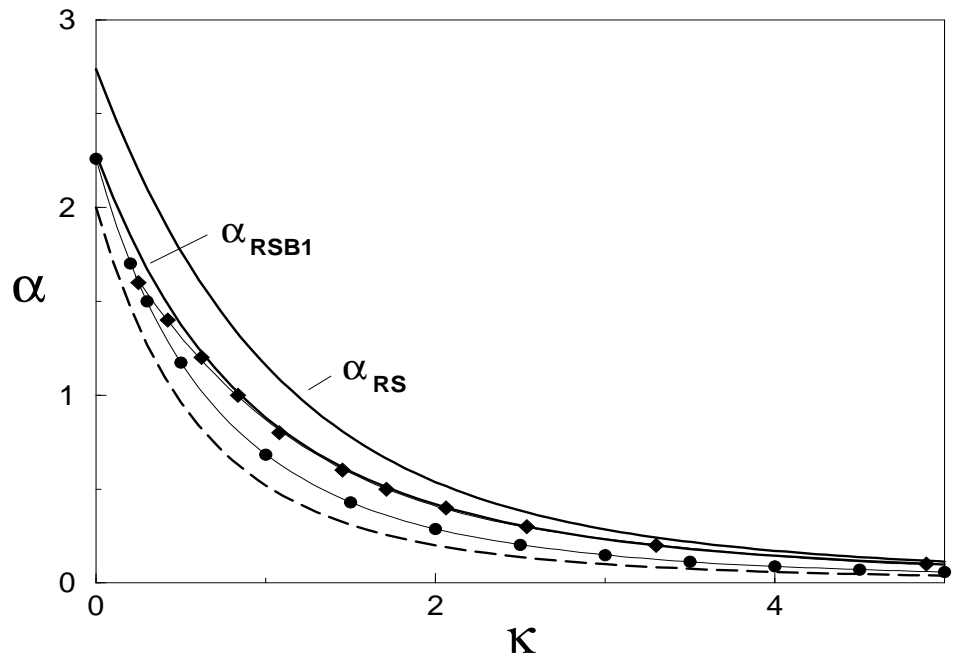}}
\caption{The maximal capacity of the coloured perceptron model II as a
function of $\kappa$. The meaning of the symbols is as in fig. 
\protect\ref{res1}.}
\label{res2}
\end{figure}

Using this learning rule we have trained networks of sizes $50 \leq N
\leq 1000$ sites (depending on the value of $\kappa$) in order to store 
perfectly as many randomly chosen patterns as possible.
For each value of $\kappa$ we have calculated the maximal capacity for  
different $N$ and extrapolated the results to $N=\infty$. Results for a
given value of $\kappa$ and $N$ are averages over 1000 samples.
As shown in figs.~\ref{res1} and \ref{res2} this algorithm performs
especially  well for small values of $\kappa$ for both the models I and II.

The second algorithm we report on is the Adatron algorithm \cite{ada}
which works in a different way. Instead of 
searching the maximal capacity for a given stability it tries to find
the maximal stability for a given capacity. The derivation of
this algorithm and a proof of its convergence are
based upon the assumption that the problem can be formulated as a quadratic
optimization with linear constraints \cite{ada,GK94}.
Such a formulation can not be given for the coloured perceptron model,
because the three different types of couplings have to be normalized  
independently and because the stability conditions
(\ref{eq:afieldxi})-(\ref{eq:afieldeta}) are more complex. Hence, a
straightforward generalization similar to the one for the Potts model
\cite{GK94} is not possible.
Below we describe a learning rule that tries to incorporate the ideas 
of the Adatron approach. We assume that the couplings can be written in
the form (cfr.,\cite{ada} and references therein)
\begin{eqnarray}
  J_i^{(1)}=\frac{1}{N}\sum_\mu x_1^\mu\xi_0^\mu\xi_i^\mu~,~~~~\nonumber
  J_i^{(2)}=\frac{1}{N}\sum_\mu x_2^\mu\eta_0^\mu\eta_i^\mu~,\\
  J_i^{(3)}=\frac{1}{N}\sum_\mu x_3^\mu\xi_0^\mu\eta_0^\mu\xi_i^\mu 
               \eta_i^\mu~,~~~~
\end{eqnarray}
where $x_r^\mu$ ($r=1,2,3$) are the so called embedding
strengths of pattern $\mu$. Then, in the case of model~II the couplings 
are updated by modifying $x_r^\mu$ with the following increments
\begin{eqnarray}
\delta x_1^\mu&=&\frac{1}{2}\max\{ -x_1^\mu-x_3^\mu,\label{x1ada}
       \gamma (1-n_1\lambda_\xi^\mu)\}~,\\\label{x2ada}
\delta x_2^\mu&=&\frac{1}{2}\max\{ -x_2^\mu-x_3^\mu,
       \gamma (1-n_2\lambda_\eta^\mu)\}~,\\\nonumber
\delta x_3^\mu&=&\frac{1}{4}(\max\{ -x_1^\mu-x_3^\mu,
         \gamma (1-n_3\lambda_\xi^\mu)\}\\
	 &+&\max\{ -x_2^\mu-x_3^\mu,\gamma (1-n_3\lambda_\eta^\mu)\})\,.
\end{eqnarray}
This is done sequentially over the patterns. We remark that again the
algorithm for model~I is somewhat different (see Appendix~B).
For each value of the
capacity we have considered system sizes $50 \leq N \leq 500$
and extrapolated the results to $N=\infty$. The best results were
obtained for a learning rate $\gamma \in (0,2)$.
Results for each size are averages over 1000 samples.
For small values of the capacity the algorithm gives better results,
both in the case of models I and II  than the first algorithm we have
discussed, as shown in figs.~\ref{res1} and \ref{res2}.  For larger
values of  the capacity, however, it performs worse. The 
results for the Adatron algorithm are displayed only in the region where
they are better than the results for the Gardner algorithm.
We remark that the numerical simulations with the different
algorithms give different results and that we have not shown their
convergence analytically such that, in principle, the values for $\alpha_c$
obtained here are lower bounds.

Looking at figs.~\ref{res1} and \ref{res2} in more detail we see that
for the whole range of $\kappa$ the values of the maximal capacity 
in model II are larger than those of a standard binary perceptron. For 
$\kappa=0$, e.g., the simulations give $\alpha_c=2.26 \pm 0.01$, which is
bigger than the maximal capacity of the binary perceptron model
\cite{Ga88} and the binary many-neuron interaction model \cite{Ko90},
both of  which have $\alpha_c=2$.
For model~I the maximal capacity at $\kappa=0$ found by simulations is
$1.78 \pm 0.01$. 

\section{Concluding remarks}

In this work we have calculated the maximal capacity per number of
couplings for two coloured perceptron models. Compared with the standard
perceptron these models have two neuronal variables per site and a local
field  that contains higher order neuron terms. The 
\noindent\rule{20.5pc}{0.1mm}\rule{0.1mm}{1.5mm}\hfill
\\\\
method
used is a generalization of the Gardner approach
and both the RS
and RSB1 results have been discussed. We expect that the latter give
very close upperbounds for the exact values.

Extensive numerical simulations  have been performed for finite systems
and extrapolated to $N=\infty$. The adaptive Gardner algorithm and the
Adatron algorithm give 
the best, but different results. Hence, the
results of the simulations can be considered only as lower bounds for
the exact maximal capacity. Additional work looking for improved algorithms 
would be welcome. 

Comparing both the RSB1 results and the results from numerical
simulations we conclude that they are in good agreement. For bigger
values of $\kappa$ they even completely coincide.
For model~I we find that at $\kappa=0$ the maximal capacity satisfies  
$1.78 \leq \alpha_c \leq1.83$. This suggests that it is equal to the maximal
capacity of the $Q=4$-Potts perceptron, i.e., $\alpha_c=1.83$ (after
appropriate rescaling of the latter \cite{GK94}). This
would parallel the situation for Hebb learning \cite{BK2}.
For model~II we have for $\kappa=0$ that $2.26 \leq \alpha_c \leq
2.28$, which is larger than the maximal capacity of the standard binary
perceptron. This is due to the fact that model~II is not a strict
mapping such that it allows for more freedom in the  determination of the 
couplings.

\acknowledgments
The authors would like to thank M.~Bouten and J. van Mourik for critical 
discussions.

\section*{Appendix A: technical details for model II}
The function $\psi_{RSB1}(\kappa,t_1,t_2,q_0^{(1)},q_0^{(3)},M)$ in 
formula (\ref{alphaRSB1}) reads

\end{multicols}
\widetext

\begin{eqnarray}
\psi_{RSB1}\left(\kappa,t_1,t_2,q_0^{(1)},q_0^{(3)},M\right)
 &=&\frac{1}{2c_1}e^{\varepsilon_3}
      \int_{-\infty}^{\frac{c1}{c}(u_1+\delta_3)}
      {\rm D}s\left[1+{\rm erf}\left(\sqrt{\frac{3r}{2c^2}} 
          \left(\frac{x_3}{\sqrt{3r}}-\delta_3
          +\frac{c}{c_1}s\right)\right)\right]
	                \nonumber\\
 &+&\frac{1}{2c_1}e^{\varepsilon_2}
       \int_{-\infty}^{\frac{c1}{c}(u_1-\delta_2)}
     {\rm D}s\left[1+{\rm erf}\left(\sqrt{\frac{3r}{2c^2}}
        \left(-\frac{x_2}{\sqrt{3r}}+\delta_2
        +\frac{c}{c_1}s\right)\right)\right]
	               \nonumber\\
  &+&\frac{1}{2c_2}e^{\phi_2}
      \int_{-\infty}^{-\frac{c2}{c}(u_1-\gamma_2)}
     {\rm D}s\left[1+{\rm erf}\left(\sqrt{\frac{3r}{2c^2}}
        \left(\frac{x_2}{\sqrt{3r}}-\gamma_2
      +\frac{c}{c_2}s\right)\right)\right]
                       \nonumber\\
  &+&\frac{1}{2c_2}e^{\phi_3}\int_{-\infty}^{-\frac{c2}{c}(u_1-\gamma_3)}
     {\rm D}s\left[1+{\rm erf}\left(\sqrt{\frac{3r}{2c^2}}
       \left(-\frac{x_3}{\sqrt{3r}}-\gamma_3
      +\frac{c}{c_2}s\right)\right)\right]
                         \nonumber\\
   &+&\frac{1}{2c'}e^{d_1}\int_{-\infty}^{-\frac{u_1}{c'}}
     {\rm D}s\left[{\rm erf}\left(\sqrt{\frac{3r}{2}}
        \left(\frac{x_3}{\sqrt{3r}}-b_1-\frac{1}{c'}s\right)\right)
     +{\rm erf}\left(\sqrt{\frac{3r}{2}}\left(-\frac{x_2}{\sqrt{3r}}-b_1-
          \frac{1}{c'}s\right)\right)\right]
	                  \nonumber\\
   &+&\frac{1}{2}\int_{-\infty}^{u_1}
     {\rm D}s\left[{\rm erf}\left(\sqrt{\frac{3r}{2}}
      \left(\frac{x_2}{\sqrt{3r}}-s\right)\right)
     +{\rm erf}\left(-\sqrt{\frac{3r}{2}}\left(\frac{x_3}{\sqrt{3r}}
              +s\right)\right)\right]\nonumber
\end{eqnarray}

\begin{multicols}{2}
\narrowtext
with D$s$ a Gaussian measure and
\begin{eqnarray}
  &c=\sqrt{1+M},~c'=\sqrt{1+M_1},~c_1=\sqrt{1+M(1+3r)},
               \nonumber\\
  &c_2=\sqrt{M_1c^2+c_1^2},~M_1=rM,~r=\frac{1-q_1'}{1-q_0^{(1)}},
                  \nonumber\\
  &x_2=\sqrt{3r}-t_2\sqrt{\frac{q_0^{(1)}}{1-q_0^{(1)}}},
         ~x_3=-\sqrt{3r}-t_2\sqrt{\frac{q_0^{(1)}}{1-q_0^{(1)}}},
	         \nonumber\\
   &\varepsilon_2=-\frac{1}{2}\frac{Mx_2^2}{c_1^2},
          ~\varepsilon_3=-\frac{1}{2}\frac{Mx_3^2}{c_1^2},
	  ~d_1=\frac{1}{2}u_1b_1
	         \nonumber\\
   &\phi_2=-\frac{1}{2c_2^2}\left(Mx_2^2(c')^2+M_1u_1^2c_1^2
          -2MM_1\sqrt{3r}u_1x_2\right),
	          \nonumber\\
   &\phi_3=-\frac{1}{2c_2^2}\left(Mx_3^2(c')^2+M_1u_1^2c_1^2
            +2MM_1\sqrt{3r}u_1x_3\right),
	            \nonumber\\
   &\delta_2=\frac{\sqrt{3r}Mx_2}{c_1^2},
       ~\delta_3=\frac{\sqrt{3r}Mx_3}{c_1^2},
        ~b_1=-\frac{M_1u_1}{(c')^2},
	               \nonumber\\
    &\gamma_2=\frac{1}{c_2^2}\left(M_1u_1c^2+M\sqrt{3r}x_2\right),
                         \nonumber
\end{eqnarray}	
\noindent\rule{20.5pc}{0.1mm}\rule{0.1mm}{1.5mm}\hfill
\begin{eqnarray}
    &\gamma_3=\frac{1}{c_2^2}\left(M_1u_1c^2-M\sqrt{3r}x_3\right),
                        \nonumber\\
   &u_1=-\frac{\sqrt{\frac{2}{3}}\kappa+\sqrt{q_1'}t_1}{\sqrt{1-q_1'}},
         ~q_1'=\frac{1}{3}q_0^{(1)}+\frac{2}{3}q_0^{(3)},
	          \nonumber\\
   &\kappa=\kappa_\xi=\kappa_\eta \,. \nonumber 
\end{eqnarray}

\section*{Appendix B: Formula for model I}
For model I the calculations are very similar. Some resulting expressions,
however, have a somewhat different structure. For completeness we write
down these expressions here. 

For the available space of couplings we get in the RS approximation 
(compare (\ref{eq:result}))
\end{multicols}
\widetext
\begin{equation}
  v=\frac{3}{2}\alpha\int\prod_r{\rm D}(s_r(q))
        \ln\left[\psi_{RS}(\kappa_\xi,\kappa_\nu,\kappa_{\xi\nu},
	s_1,s_2,s_3,q)\right]
      -\frac{3}{2}\alpha\ln 4+\frac{3}{2}\left(\ln(1-q)+\frac{q}{1-q}+
      \ln 2\pi\right)
           \nonumber
\end{equation}
with
\begin{eqnarray}
\psi_{RS}(\kappa_\xi,\kappa_\nu,\kappa_{\xi\nu},s_1,s_2,s_3,q)
&=&\left(\int_{-\infty}^{l_1}{\rm d}u_1
             \int_{l_4}^{\infty}{\rm d}u_2
	                   \int_{l_5}^{\infty}{\rm d}u_3
      +\int_{-\infty}^{l_2}{\rm d}u_2
              \int_{l_6}^{\infty}{\rm d}u_1
	               \int_{l_7}^{\infty}{\rm d}u_3
\right.\nonumber\\  
  &+&\left.\int_{-\infty}^{l_3}{\rm d}u_3
          \int_{l_8}^{\infty}{\rm d}u_1\int_{l_9}^{\infty}{\rm d}u_2
     +\int_{l_1}^{\infty}{\rm d}u_1\int_{l_2}^{\infty}{\rm d}u_2
              \int_{l_3}^{\infty}{\rm d}u_3\right)
     \prod_r\frac{e^{-\frac{1}{2}u_r^2}}{\sqrt{2\pi}}
\end{eqnarray}

\begin{multicols}{2}
\narrowtext
where
\begin{eqnarray}
  l_i&=&\frac{L_i+s_i}{\sqrt{1-q}},~~~i=1,2,3,~~
            \nonumber\\
  l_4&=&\frac{L_1+L_2+s_1+s_2}{\sqrt{1-q}}-u_1,~l_6=l_4+u_1-u_2,
           \nonumber\\\nonumber
  l_5&=&\frac{L_1+L_3+s_1+s_3}{\sqrt{1-q}}-u_1,~l_8=l_5+u_1-u_3,
           \nonumber\\
  l_7&=&\frac{L_2+L_3+s_2+s_3}{\sqrt{1-q}}-u_2,~l_9=l_7+u_2-u_3,
           \nonumber\\
  L_1&=&\frac{1}{2}(\kappa_\xi-\kappa_\eta+\kappa_{\xi\eta}),~
  L_2=\frac{1}{2}(-\kappa_\xi+\kappa_\eta+\kappa_{\xi\eta}),
            \nonumber\\
   L_3&=&\frac{1}{2}(\kappa_\xi+\kappa_\eta-\kappa_{\xi\eta})\nonumber
\end{eqnarray}
and $q$ taking those values that minimizes $v$.
Thus, for $\kappa=\kappa_\xi=\kappa_\eta=\kappa_{\xi\eta}$ the
maximal capacity in the RS approximation can be written as
\begin{eqnarray}
\alpha_{RS}(\kappa)=\hspace{6.5cm}\nonumber\\
\lim_{q\rightarrow 1}
   \left\{\frac{-\ln(1-q)-\frac{q}{1-q}-\ln 2\pi}
              {\int\prod_r{\rm D}(s_r(q))
    \psi_{RS}(\kappa,\kappa,\kappa,s_1,s_2,s_3,q)-\ln 4}\right\}\nonumber
\end{eqnarray}

For the RSB1 approximation with the form of the order parameters given
by (\ref{rsb1a}) the maximal capacity reads
\begin{eqnarray}
   \alpha_{RSB1}(\kappa)=\hspace{6.5cm}\nonumber
\end{eqnarray}
\noindent\hfill\rule[-1.5mm]{0.1mm}{1.5mm}\rule{20.5pc}{0.1mm}
\begin{eqnarray}   
   \min_{q_0,M}\left\{\frac{-\ln(1+M)
                              -\frac{q_0M}{(1+M)(1-q_0)}}
              {\int\prod_r{\rm D}t_r\ln\psi_{RSB1}
     \left(\kappa,t_1,t_2,t_3,q_0,M\right)}\right\} \hspace{1.5cm}\nonumber
\end{eqnarray}
with $\psi_{RSB1}(\kappa,t_1,t_2,t_3,q_0,M)$ a linear combination of 
thirty-four, mostly double, integrals
over error functions. An interested reader can find a complete
formula for $\psi_{RSB1}(\kappa,t_1,t_2,t_3,q_0,M)$ in \cite{PhD}.

Finally, the learning algorithms for model~I differ in the way that the 
couplings $J^{(1)}$ and $J^{(2)}$ are updated. We have for the adaptive
Gardner algorithm
\begin{eqnarray}
  J_i^{(1)}&\rightarrow&J_i^{(1)}+\xi_0^{\mu}\xi_i^{\mu}\frac{1}{2}
            \left[\left(\kappa_\xi-\lambda_\xi^\mu\right)
         \Theta\left(\kappa_\xi-\lambda_\xi^\mu\right)\right.
	          \nonumber\\		   
      &+&\left(\kappa_{\xi\eta}-\lambda_{\xi\eta}^\mu\right)\left.
     \Theta\left(\kappa_{\xi\eta}-\lambda_{\xi\eta}^\mu\right)\right]
               \nonumber\\
  J_i^{(2)}&\rightarrow&J_i^{(2)}+\eta_0^{\mu}\eta_i^{\mu}\frac{1}{2}
            \left[\left(\kappa_\eta-\lambda_\eta^\mu\right)\right.
         \Theta\left(\kappa_\eta-\lambda_\eta^\mu\right)
	         \nonumber\\		   
     &+&\left(\kappa_{\xi\eta}-\lambda_{\xi\eta}^\mu\right)\left.
          \Theta\left(\kappa_{\xi\eta}-\lambda_{\xi\eta}^\mu\right)\right]
	  \nonumber
\end{eqnarray}
instead of (\ref{j1per}) and (\ref{j2per}) and for the Adatron algorithm
we take
\begin{eqnarray}
\delta x_1^\mu=
    \frac{1}{2}(\max\{ -x_1^\mu-x_3^\mu,\gamma (1-n_1\lambda_\xi^\mu)\}
              \nonumber\\
    +\max\{ -x_1^\mu-x_2^\mu, \gamma (1-n_1\lambda_{\xi\eta}^\mu)\})
              \nonumber\\
\delta x_2^\mu=\frac{1}{2}(\max\{ -x_2^\mu-x_3^\mu, 
                      \gamma (1-n_2\lambda_\eta^\mu)\}
               \nonumber\\
    +\max\{ -x_1^\mu-x_2^\mu,\gamma (1-n_2\lambda_{\xi\eta}^\mu)\})
            \nonumber
\end{eqnarray}
instead of (\ref{x1ada}) and (\ref{x2ada}).

\end{multicols}
\widetext

\end{document}